\documentclass[adp,fleqn]{w-art}

\usepackage{times}
\usepackage{w-thm}
\usepackage[]{graphicx}
\chardef\bslash=`\\ 

\usepackage{amsmath,amsfonts,amssymb,bm,psfrag}
\renewcommand{\vec}[1]{{\bf #1} }
\newcommand{\PP}{\tilde{P}\,}
\newcommand{\RE}{\text{Re}\,}
\newcommand{\IM}{\text{Im}\,}

\newcommand{\eps}{\varepsilon}

\newcommand{\vecG}{{\bm G}}

\newcommand{\md}{\text{d}}
\newcommand{\me}{\text{e}}
\newcommand{\mi}{\text{i}}
\newcommand{\sgn}{\,\text{sgn}\,}

\newcommand{\w}{\omega}

\begin{document}
\DOIsuffix{theDOIsuffix}
\Volume{12}
\Issue{1}
\Copyrightissue{01}
\Month{01}
\Year{2003}
\pagespan{1}{}



\title[Optical Conductivity of Clean Metals]{Optical Conductivity of Clean Metals}


\author[A.\ Rosch]{A. Rosch\footnote{Corresponding
     author \quad E-mail: {\sf rosch@thp.physik.uni-koeln.de}, Phone: +49\,221\,470\,4994,
     Fax: +49\,221\,470\,2189}}
\address[]{Institut f\"ur Theoretische Physik, Universit\"at zu K\"oln, D-50937 Cologne, Germany.}
\dedicatory{Dedicated to the memory of Paul Drude.}

\begin{abstract}
 We briefly review some basic aspects of transport in clean metals focusing on the role of
 electron-electron interactions and neglecting the effects
  of impurities, phonons and interband transitions.
 Both for small Fermi surfaces of two and three-dimensional metals and open Fermi surfaces of quasi
 one-dimensional metals the dc conductivity $\sigma$ is largely dominated by
 momentum and pseudo-momentum conservation, respectively. In general, the frequency and temperature
  dependencies of $\sigma(\w,T)$ have very little in common.  For small Fermi
  surfaces in three dimensions we find for example that the scattering
  rate is quadratic in frequency,
  $\Gamma \propto \w^2$, even in the {\em absence} of a $T^2$ contribution.
\end{abstract}
\maketitle





\section{Introduction}
More than a century after Drude  \cite{drude} published his
pioneering work 'Zur Ionentheorie der Metalle' ('On the ion theory
of metals'), the optical conductivity is one of the most useful
tools \cite{dressel} to investigate the basic properties of metals.
Both the excitation spectrum of metals (gaps, phonons, magnons,
interband transitions, ...) and the scattering mechanisms leave
their distinct traces in transport. It is surprising that the Drude
model, which was developed before even the concepts of  electrons
(and Fermions in general) had been established,  describes much of
the phenomenology of low-frequency transport in metals correctly.

In this paper we will review some basic (but apparently not very
well known) facts about the conductivity and low-frequency optical
properties of clean metals. We will concentrate our discussion on
the lowest frequencies, i.e. on the Drude peak at low but finite
temperature, and on the $T=0$ optical conductivity at the lowest
frequencies. We will not consider the effects of disorder or phonons
but study only the effects of electron-electron interactions in
clean Fermi liquids. Parts of results discussed here have already
been published in Ref.~\cite{dWave} ($T=0$ optical conductivity of
d-wave superconductors and metals) and Ref.~\cite{peter} (clean
quasi one-dimensional Fermi liquids at low $T$).

In a Galilean invariant system, the electrical conductivity is
infinite as the current $\bm J= \bm P/m$ is conserved, where $\bm P$
is the momentum and $m$ the electron mass. In a crystal,
translational invariance is partially broken which has two
consequences. First, momentum is not conserved anymore as it can be
transferred to the underlying lattice in units of the reciprocal
lattice vectors $\bm G_i$ (using $\hbar=1$) via so-called ''Umklapp
scattering processes''. A macroscopic momentum will therefore decay.
Second, due to band-structure effects, current is not anymore
proportional to the momentum. Even processes which conserve momentum
can induce a partial decay of the current. As we elaborate below,
non-Umklapp processes will lead to a decay of current components
perpendicular to the momentum while the parallel component is not
affected. The rather different role of current and momentum is
reflected in the disparate frequency and temperature dependence of
the optical conductivity. While the $T$ dependence of the dc
conductivity is often determined by the decay rate of the momentum
(or of the pseudo-momenta introduced in Sec.\ref{momenta}), the
finite frequency properties reflect that $\bm J \neq \bm P/m$, as we
will explain in detail below.

In the following section we will first explore the role of momentum
conservation. We emphasize that momentum has no unique definition in
lattice and that the concept of momentum can and should be
generalized e.g. in quasi one-dimensional systems or d-wave
superconductors. We will first investigate temperature and then the
frequency dependence of the optical conductivity.

\section{Momenta and pseudo-momenta in lattices}\label{momenta}
In a one-band model, the electrical current  is given by
$\vec{J}=\sum_{\sigma,\vec{k} \in BZ} \vec{v}_{\vec{k}}
c_{\vec{k}\sigma}^\dagger c_{\vec{k}\sigma}$ where the momentum sum
extends over the first Brillouin zone (BZ) and $\vec{v}_{\vec{k}}=d
\epsilon_{\vec{k}}/d \vec{k}$ is the velocity of electrons with
creation operator $c_{\vec{k}\sigma}^\dagger$ and band energy
$\epsilon_{\vec{k}}$. While in single band models the current can
simply be obtained  from a continuity equation, this is not the case
in more complex multi-band models where inter-band matrix elements
of the current depend on the details of the underlying microscopic
model. As we are only interested in the conductivity at low
frequencies and temperatures, we will not discuss such inter-band
transitions in the following.

\begin{figure}
a)\ \includegraphics[width=0.28 \linewidth]{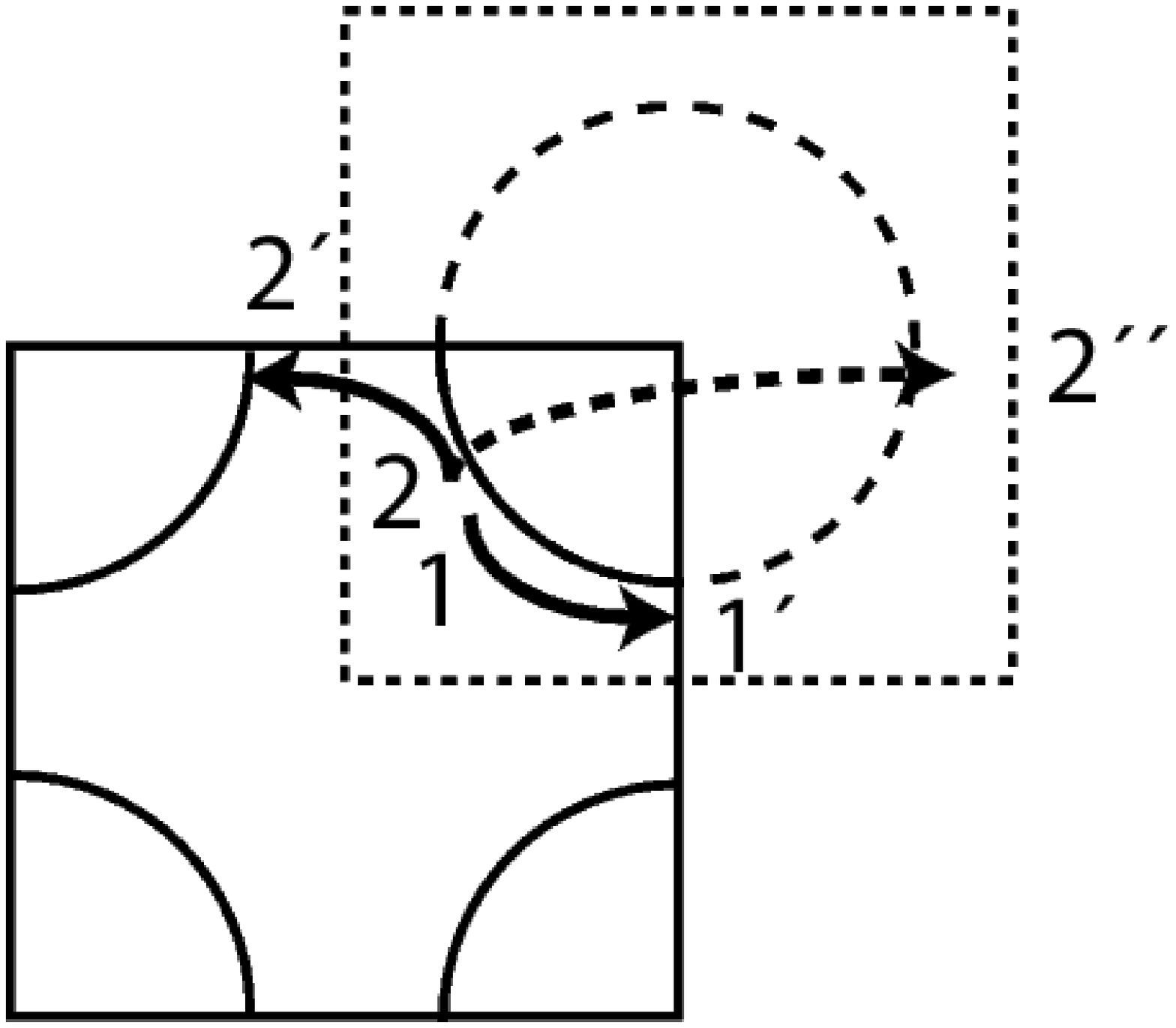}\ b)
\includegraphics[width=0.28 \linewidth]{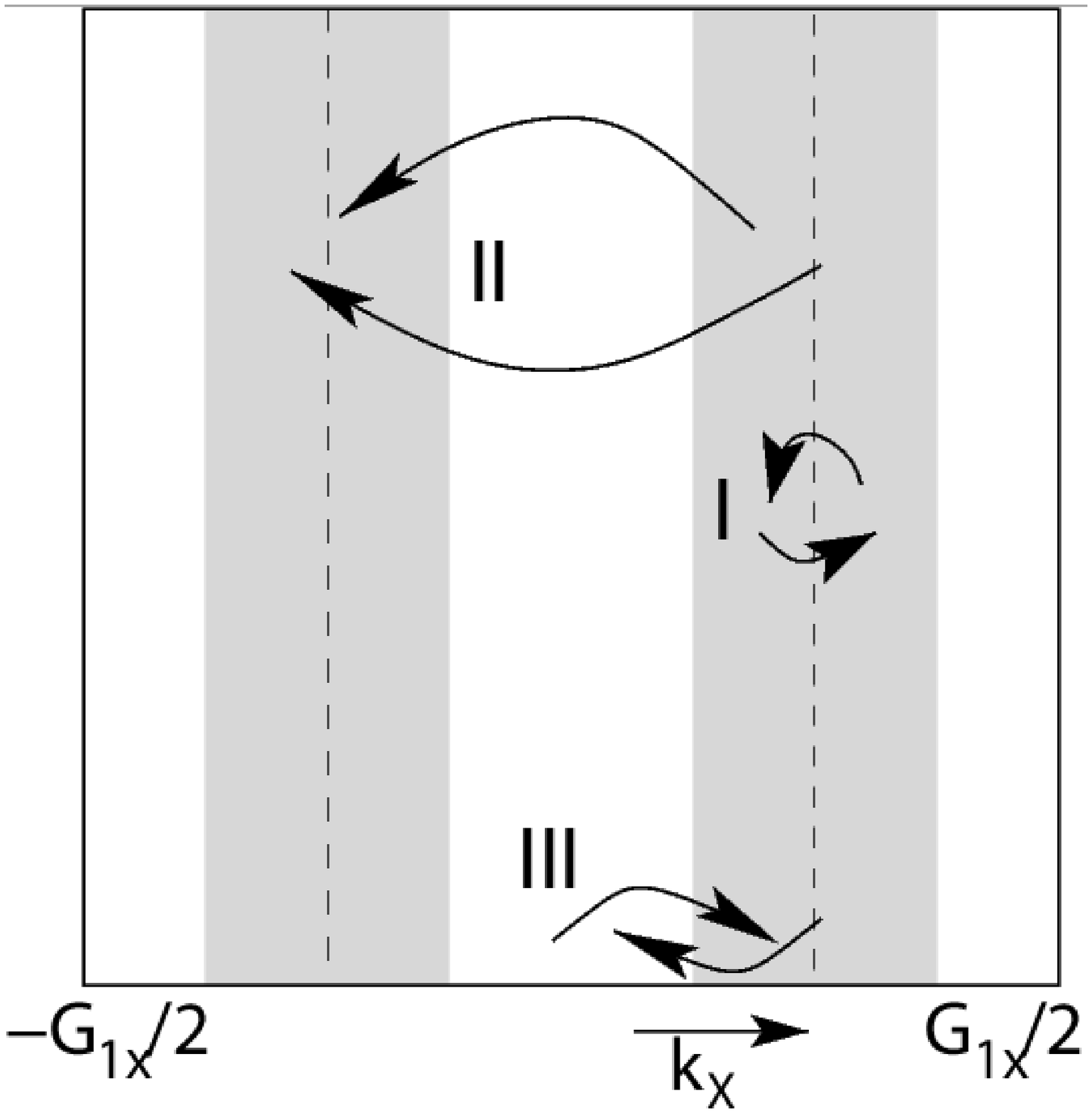}\ c)
\includegraphics[width=0.32 \linewidth]{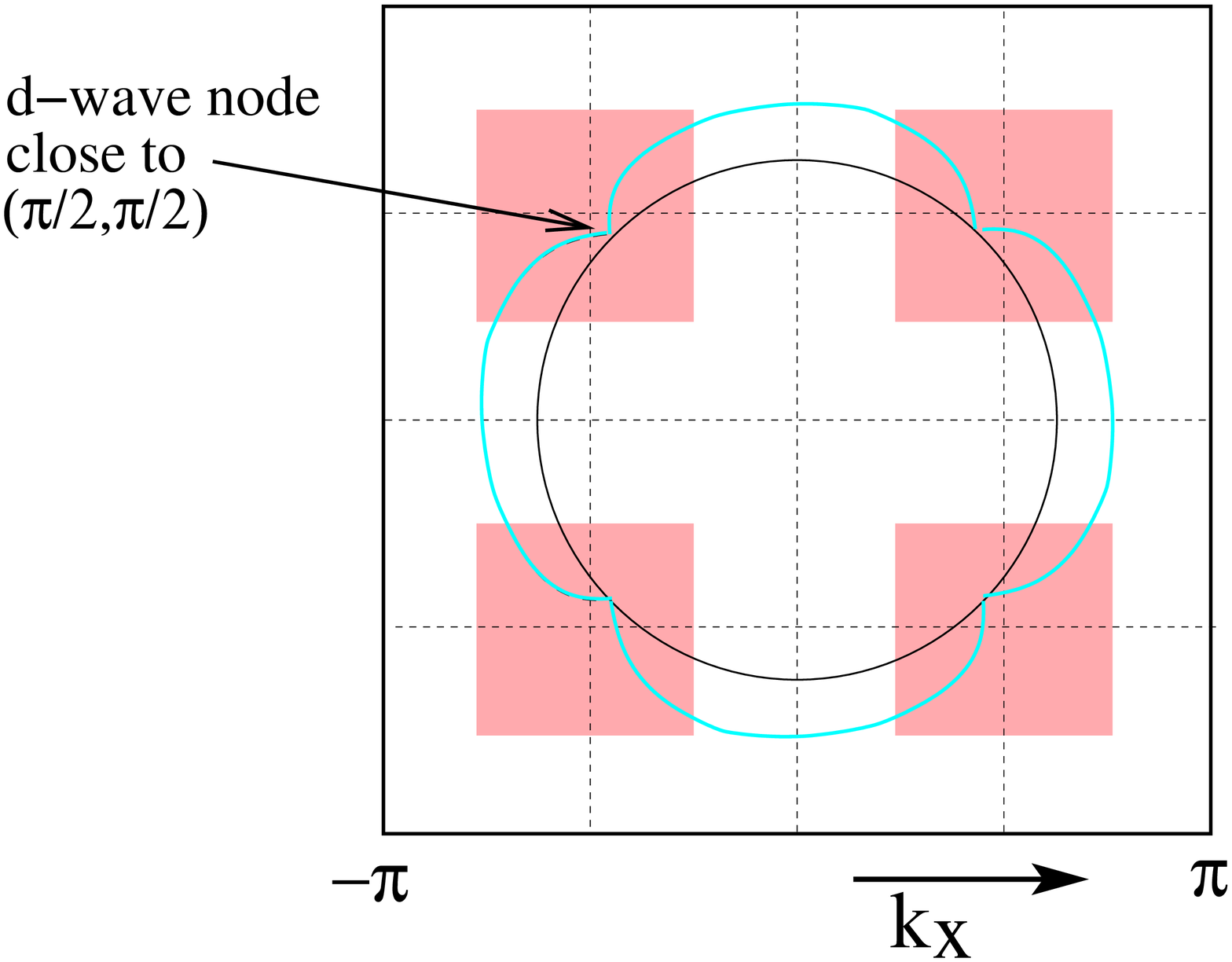} \caption{\label{fig1}
a) Sketch of a scattering process using two different Brillouin
zones. Two particles with state 1 and 2 are scattered into the
states 1' and 2' (or equivalently 2''). While in hole-like Fermi
surface (solid lines) no momentum is transferred to the lattice, the
same scattering event in the dashed unit cell (replacing 2' by the
equivalent 2'')  is an Umklapp process. b)
  Fermi surface  of an anisotropic metal close to half filling.  Both
  ``forward'' (I) and ``Umklapp'' (II) scattering processes do not
  lead to a decay of the pseudo-momentum $\PP_{21}$ as long as the
  momenta are within the shaded area. The scattering event III
   leads to a decay of $\PP_{21}$.  c) Also in a d-wave superconductor with point nodes close to
   $(\pi/2,\pi/2)$, the pseudo momentum $\PP_{21}$ is the slowest
   mode in the systems as two-particle scattering events cannot relax $\PP_{21}$ if all momenta remain within
    the shaded
   regions.}\end{figure} As the velocity is a periodic function of momentum,
$\vec{v}_{\vec{k}+\bm G_i}=\vec{v}_{\bm k}$, the definition of $\bm
J$ is obviously independent of the choice of the unit-cell in
reciprocal space. This is, however, not the case for the momentum
$\bm P =\sum_{\sigma,\vec{k} \in BZ} {\vec{k}}
c_{\vec{k}\sigma}^\dagger c_{\vec{k}\sigma}$ which depends on the
chosen (first) Brillouin zone. For a given definition of $\bm P$,
 one can separate all scattering
events in 'non-Umklapp' and 'Umklapp' processes, where the latter
are defined by a change of the momentum by a reciprocal lattice
vector. The choice of the BZ will consequently decide whether a
given scattering event is an Umklapp process as is shown in
Fig.~\ref{fig1}a. As it is common knowledge that in clean systems
only Umklapp processes contribute to the dc conductivity and as the
physics should not depend on the completely arbitrary choice of the
BZ, one has to ask which definition of the momentum operator is the
'correct' one. Momentum conservation is only important in situations
where $\bm P$ decays very slowly. Therefore the relevant momentum is
the mode with the slowest decay rate. This can for example be seen
within a simple Boltzmann equation (essentially equivalent argument
hold also in situations, where a Boltzmann description is not
possible, see e.g. \cite{roschPRL}). After linearization, the
Boltzmann equation in the presence of an electric field $\bm E$
takes (in units where $e=1$) the form
\begin{eqnarray}
\partial_t  \Phi_{\bm k}+\bm v_{\bm k} \bm E &=& \sum_{\bm k'} M_{\bm k \bm k'} \Phi_{\bm k'}\\
\sigma_{ii}(\w) &=& -\sum_{\bm k,\bm k'} v^i_{\bm k} [M- i
\w]^{-1}_{\bm k,\bm k'}v^i_{\bm k'}\frac{\partial f_0}{\partial
\epsilon_{\bm k}}\label{sigma}
\end{eqnarray}
where $\sigma_{ij}(\w)$ is the (diagonal) optical conductivity
tensor, $f_{\bm k}=f_0(\epsilon_{\bm k})-\Phi_{\bm k}
\partial f_0/\partial \epsilon_{\bm k}$ the distribution function
linearized around the Fermi function $f_0$, and $M_{\bm k \bm k'}$
the well-known \cite{ziman} scattering matrix. For a diagonal
scattering matrix $M=\frac{1}{\tau} \delta_{\bm k \bm k'}$ one
recovers the celebrated Drude formula. For the models considered in
this paper, such a relaxation-time approximation cannot be applied
as certain eigenvectors of $M$ have very small eigenvalues.
Physicswise, this is a consequence of the approximate conservation
of momentum or pseudo momentum (see below).

Let us first consider the well-known case of a Fermi liquid with a
small Fermi surface, where Umklapp processes are strongly
suppressed. In this case the inverse of the matrix $M$ is dominated
by a single eigenvector with a very small eigenvalue. This
eigenvector is with high precision the momentum parallel to the
current. Introducing the scalar product $(a|b)=-\sum_{\bm k} a_{\bm
k} b_{\bm k} \partial f_0/\partial \epsilon_{\bm k}$, one finds in
such a situation~\cite{ziman}
\begin{equation}
\sigma_{xx}(\w=0) = (v^x | M^{-1} | v^x) =\max_\Phi
\frac{(v^x|\Phi)(\Phi|v^x) }{ (\Phi | M | \Phi)} \approx
\frac{(v^x|p^x)(p^x|v^x) }{ (p^x | M | p^x)} \label{mm}
\end{equation}
where the variational vector $\Phi$ takes at the maximum a value
proportional to $M^{-1} |v^x)$.

We now use (\ref{mm}) to generalize the concept of momentum and to
resolve the above discussed ambiguities related to the choice of the
BZ. First of all, momentum conservation (modulo reciprocal lattice
vectors) is only relevant if the momentum decays much slower than
other eigenvectors of $M$. For example,  for a complex
three-dimensional Fermi surface both Umklapp and non-Umklapp
scattering rates are proportional to $T^2$  and momentum
conservation and its generalizations are therefore not important.
Second, when defining the momentum, $P_x =\sum_{\sigma,\vec{k} \in
BZ} k_x c_{\vec{k}\sigma}^\dagger c_{\vec{k}\sigma}$, one should use
the BZ which results in the {\em slowest} decay rate of $P_x$ or,
equivalently, in the largest value on the RHS of Eq.~(\ref{mm}).
Third, in many situations it is necessary to generalize the concept
of momentum by introducing so-called 'pseudo momenta'
\cite{roschPRL} as the relevant eigenvector of $M$ is not
proportional to momentum defined for any choice of the BZ.

Consider an arbitrary one-particle operator $\PP=\sum_{\bm k}
\tilde{p}_{\bm k}c_{\vec{k}\sigma}^\dagger c_{\vec{k},\sigma}$ with
the time derivative $\partial_t \PP=\mi \sum_{\vec{k}
\vec{k}'\vec{q}\sigma\sigma'}U_\vec{q}
(\tilde{p}_{\vec{k}}+\tilde{p}_{\vec{k}'}-\tilde{p}_{\vec{k}+\vec{q}}-
\tilde{p}_{\vec{k}'-\vec{q}})c_{\vec{k}\sigma}^\dagger
c_{\vec{k}+\vec{q},\sigma}c_{\vec{k}'\sigma'}^\dagger
c_{\vec{k}'-\vec{q},\sigma'} $ in the presence of an arbitrary
density--density interaction $U_\vec{q}$. We want to construct an
operator $\PP$ which (i) does not decay by 2-particle processes
close to the Fermi surface and (ii) has a sizeable overlap  [defined
by $(\tilde{p}|v_x)]$ with the current in $x$ direction. We
therefore require that $\tilde{p}_{\vec{k}}$ is finite close to the
Fermi surface while
$\tilde{p}_{\vec{k}}+\tilde{p}_{\vec{k}'}-\tilde{p}_{\vec{k}+\vec{q}}-
\tilde{p}_{\vec{k}'-\vec{q}}$ should vanish for low-energy
excitations. For a small Fermi surface, the usual momentum has these
properties. If we are able to construct such a quantity in more
complicated situations, we will call $\PP$ 'pseudo momentum'.

Such a construction is indeed possible when the low-energy
excitations cover only a small part of the BZ as it is the case e.g.
for quasi one-dimensional metals and unconventional superconductors
with point or line nodes. For illustration, let us define a set of
pseudo momenta labeled by the two integers $n$ and $m$
(generalizations are possible)
\begin{eqnarray}\label{pseudo}
\tilde{P}_{n m}=\sum_{\vec{k}} \delta k_{nm} c^\dagger_{\vec{k}}
c_{\vec{k}}=\sum_{\vec{k}} (k_x-\frac{m}{n}\frac{G_x}{2} \sgn k_x)
c^\dagger_{\vec{k}} c_{\vec{k}}
\end{eqnarray}
where $G_x$ is the $x$ component of a reciprocal lattice vector. For
$m=0$ one recovers the standard momentum. The sign-function $\sgn
k_x$ implies that for positive/negative $k_x$ the momentum is
measured with respect to the line $k_x=\pm
\frac{m}{n}\frac{G^1_x}{2}$, i.e. the dashed lines in
Figs~\ref{fig1}b and c. These pseudo momenta dominate transport if
the relevant low-energy excitations are located close to these lines
as can be the case in quasi one-dimensional metals or e.g. a
two-dimensional d-wave superconductor with nodes close to
$(\pi/2,\pi/2)$.

For the quasi one-dimensional Fermi surface shown in
Fig.~\ref{fig1}b, two types of scattering processes dominate for low
$T$. 'Forward' processes of type I conserve momentum and therefore
also the pseudo-momentum $\PP_{21}$ with $\delta
k_{21}=k_x-\sgn(k_x) G_x/4$. Umklapp processes of type II transfer
the momentum $G_x$ to the lattice: momentum decays rapidly. But as
two particles move from the right to the left Fermi surface the
pseudo momentum gets an extra contribution $2 \cdot 2 \cdot
G_x/4=G_x$ and remains conserved. More precisely, $\PP$ does not
decay by two-particle processes, if all four momenta are in the
shaded region of Fig.~\ref{fig1}, i.e. $|\delta k_{21}|< G_{1x}/4$.
While the momentum is conserved modulo reciprocal lattice vectors,
$G$, the pseudo momentum $\tilde{P}_{nm}$ can decay in quanta of
size $G_x/n$, for example by process
 III shown in Fig.~\ref{fig1}. Such
 scattering processes are exponentially suppressed even at moderate
temperatures. We conclude that in cases where the low-energy
excitations are located in the shaded areas of Fig. \ref{fig1}b and
c, the Boltzmann equation predicts for finite $(v^x|\tilde{p})$ an
{\em exponentially large} dc-conductivity which has its origin in
the approximate conservation of the pseudo-momentum $\PP_{21}$.

The overlap of electrical current and pseudo momentum
$(v^x|\tilde{p})$, see Eq.~(\ref{mm}), is easily calculated within
the approximation scheme underlying the Boltzmann approach. By
partial integration one obtains
\begin{equation}
(v^x|\tilde{p}_{nm})=\sum_{\bm k} f^0_{\bm k}
\partial_{k_x} \delta k_{nm}=\Delta \rho=\rho-\rho_{\rm max}n/m
\end{equation}
 where $\rho$ is the total particle density, $\rho_{\rm max}$ the
density of a filled band. For an exactly half-filled band one
therefore has $(v^x|\tilde{p})=0$ and conservation of $\PP_{21}$ is
{\em not} important but dominates transport for small doping away
from half-filling. In Ref.~\cite{peter} it is shown that these
statements also hold beyond perturbation theory even if a simple
Fermi liquid description is not possible.

\begin{figure}
\includegraphics[width=0.69 \linewidth]{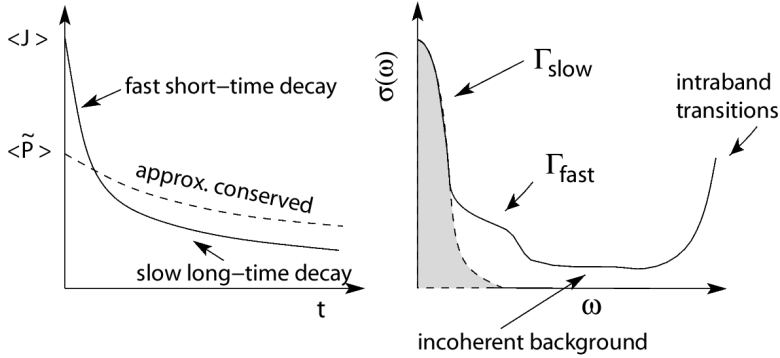}
 \caption{ A finite current
  $\langle J \rangle$ typically decays initially very rapidly. However, if a
  (pseudo-) momentum $\PP$ is present, which is approximately conserved,
the component of the current parallel to $\PP$ will decay much
slower, giving rise to a
  low-frequency peak in the optical conductivity with a width
   which is determined by the slow decay rate $\Gamma_\text{slow} \ll
   T^2$. The interplay of slow and fast decay, $\Gamma_\text{fast}\sim
   T^2$,  results in a non-Drude like frequency dependence of the
   low-$\w$
   peak. For $\w \gg \Gamma_{\text{slow}}$, a constant incoherent
   background should show up in generic three-dimensional materials. Higher
   frequencies are usually dominated by intraband transitions (and
   phonons etc.).
\label{fig2} }
\end{figure}
How does the presence of an approximately conserved pseudo-momentum
affect the optical conductivity (a more quantitative version of the
following arguments can be found in Ref. \cite{peter}, a numerical
test in Ref.~\cite{garst})? Consider the following Gedanken
experiment illustrated in Fig.~\ref{fig2}: prepare a state with a
finite current $\langle J(t=0) \rangle>0$ and switch off the driving
electric field at time $t=0$.  As the current is not conserved, it
will decay rather fast with a typical rate which we denote by
$\Gamma_{J}$. The initial state with finite current will typically
also have a finite pseudo momentum [assuming $(v^x|\tilde{p})\neq
0$]  which will decay at a much slower rate than the current;
$\Gamma_{\PP}\ll \Gamma_{J}$. The slow decay of the pseudo-momentum
will, as time goes on, induce a slow decay of the current since
 a state with finite pseudo-momentum will typically carry a
finite current as $(v^x|\tilde{p})\neq 0$ according to our
assumptions. Thus, in the long time limit, a finite fraction,
$D/D_0$, of $\langle J \rangle$ will not decay with the fast rate
$\Gamma_{J}$ but with the much smaller rate $\Gamma_{\PP}$ as is
shown schematically in Fig.~\ref{fig2} (the effect of exact
conservation laws has been studied a long time ago by Mazur and
Suzuki \cite{mazur,suzuki} and more recently by Zotos and coworkers
\cite{zotos}) . The slow long-time decay of some fraction of the
current leads to a corresponding long-time tail in the (equilibrium)
current-current correlation function. Therefore, one expects a
low-frequency peak in the optical conductivity which carries the
fraction $D/D_0$ of the total weight $D_0$ (a calculation of $D$
within Fermi liquid theory is given in Ref.~\cite{peter}). The width
of the low-frequency peak is determined by the decay-rate of $\PP$.
As the weight of a peak is approximately given by its width
multiplied with its height, we expect that the height and therefore
the dc-conductivity is of the order of $D/\Gamma_{\PP}$.

\section{Temperature dependence of the dc conductivity:\\
Small Fermi surfaces and quasi one-dimensional metals}\label{sec2}

The qualitative analysis of the Boltzmann equation sketched above
suggests that the dc conductivity of a clean metal is  {\em
exponentially} large \cite{peierls,landau}, $\sigma \sim (\Delta
\rho)^2 e^{\Delta E/T}/T^2$, if the Fermi surface is either
sufficiently small or sufficiently one-dimensional such that a
momentum or pseudo-momentum can be found which does not decay by
two-particle scattering processes close to the Fermi surface. Here,
the so-called Umklapp gap $\Delta E$ depends on the detailed shape
of the Fermi surface.

However, it is not difficult to see that the exponential $T$
dependence of $\sigma$ (an often cited result going back to Peierls
\cite{peierls,landau}) is an artifact of a Boltzmann equation which
considers only two-particle scattering processes. A low-energy
scattering process involving $N$ particles can -- for sufficiently
large $N$ -- transfer momentum to the lattice even for small Fermi
surfaces. We therefore have to generalize our discussion to $N$
particle scattering events. Due to conservation of lattice momentum
and particle number, $ \tilde{P}_{nm}$ can be altered only by an
amount in quanta of size $\frac{1}{n} G^1_x$ (for small Fermi
surfaces use $m=0$ and $n=1$ such that $\delta k_{nm}=k$).
Therefore, a relaxation of $\PP$ is {\em not} possible if all $2N$
  pseudo momenta $\delta k_{ix}$ involved in an $N$-particle scattering process
  are smaller than $\frac{1}{2 N}\frac{1}{n} G^1_x$.
 For a given Fermi surface, the decay of $\PP$
  by $N$-particle collision  at low $T$ is possible only  for
\begin{eqnarray}\label{minN}
N>\frac{G_{1x}/(2 n)}{ \max |\delta k^F_{nm}|},
\end{eqnarray}
where $\max |\delta k^F_{nm}|$  is the maximal distance of the Fermi
surface from the plane $k_x=\pm  \frac{m}{2 n} G_{1x}$ (dashed line
in Fig.~\ref{fig1}b). In the case of a small Fermi surface, $\delta
k^F_{nm}$ is just the Fermi momentum.

The $T$ dependence of the decay rate of the (pseudo-) momentum
arising from an $N$ particle scattering event can be obtained from
the usual phase space arguments: when a particle of energy $\sim T$
decays into $2N-1$ particle and hole excitations, one of the
energies is fixed by energy conservation, and the remaining $2 N-2$
energies each have a phase-space of order $T$. Therefore,
$\Gamma_{\PP} \propto T^{2 N-2}$
\begin{eqnarray}
\sigma(\w = 0) \sim \frac{(\Delta \rho)^2}{T^{2 N-2}} \label{sigmaP}
\end{eqnarray}
where the integer $N$ is the smallest value consistent with
(\ref{minN})  \cite{new,remark}. The prefactor of $\Gamma_{\PP}$ and
therefore also the prefactor in (\ref{sigmaP}) depends in a rather
delicate way on the strength and range of the interaction, the
screening, and the band-curvature, and the prefactor in Eq.
(\ref{sigmaP}) will be very large for a weakly interacting system
and large $N$.
 The crossover from a high-temperature
regime where 2-particle scattering dominates to low temperatures can
crudely be mimicked by an interpolating formula of the type $
\Gamma_{\PP} \sim U^2 T^2 e^{-\Delta E/T}+ U^N T^{2 N-2}$, where $U$
is an effective interaction and $\Delta E$ the Umklapp gap for
2-particle collisions.

We want to stress that the analysis given above is valid both for
small Fermi surfaces and quasi one-dimensional systems with moderate
curvature of the Fermi surface. It holds for interactions of
arbitrary strength as long as a Fermi liquid description is
possible. For practical purposes, a very clean system is needed to
observe a large exponent $N$ as otherwise the (pseudo) momentum will
decay by impurity scattering. In this case, one recovers the usual
$T^2$ correction to the resistivity for the lowest $T$, however with
a prefactor which depends on the amount of disorder in the system.

\section{Frequency dependence of the optical conductivity}

In the previous section, we have discussed the $T$ dependence of the
conductivity for $\w=0$ or more precisely $\w \ll \Gamma_{\PP}$
where $\Gamma_{\PP}$ is the decay rate of the momentum at $\w=0$. In
this section, we consider the small-frequency dependence of the
$T=0$ conductivity, or more precisely the limit $\w \gg \Gamma_J$,
where $\Gamma_J$ is the $\w=0$ decay rate of the {\em current}. We
do not investigate the intermediate regime $\Gamma_{\PP} \lesssim \w
\lesssim \Gamma_J$. For simplicity, we also concentrate our
discussion on systems with small Fermi surfaces to avoid the
complications arising from the numerous crossover scales in the
quasi one-dimensional system. In the limit $\w \to 0$, $T=0$, the
frequency dependence of the optical conductivity of a quasi
one-dimensional system with finite curvatures of the Fermi sheets is
characterized by the same power laws as a generic three-dimensional
metal.

 For $\w>0$, $T=0$ and in the absence
of disorder, it is possible to calculated the  optical conductivity
from a straightforward perturbative evaluation of the Kubo formula.
\begin{equation}
  \RE \sigma(\omega) = \frac{1}{\w} \, \IM
  \left \langle\!\left\langle J , J \right \rangle\!\right\rangle_\w
  =-\frac{\mi}{\w}
\int_0^\infty
  \md t \,\me^{\mi (\w+\mi 0) t} \langle[J(t),
  J(0)]\rangle.
\end{equation}
In contrast, the calculation of dc transport at finite $T$ to
leading order in perturbation theory involves usually the solution
of some integral equations (for example, the quantum Boltzmann
equation or equivalently a vertex equation) or equivalently a matrix
inversion, see Eq. (\ref{sigma}). The simplifications arising at
$T=0$, $\w>0$ can be seen by inspecting the (generalized) Drude
formula
\begin{equation}\label{sig} \sigma(\w)=\frac{\chi}{\Gamma(\w)-i
\omega} \end{equation}
 where $\chi$ is identified with the total
optical weight and $\RE \Gamma(\w)$
 is the (frequency-dependent) scattering rate. For $\w=0$ and $T>0$ the conductivity is a singular function
 of $\Gamma(\w=0)$ and therefore an infinite resummation of perturbation
  theory [e.g. by solving a Boltzmann equation or by inverting the matrix in
  Eq.~(\ref{sigma})]
 is required to calculate $\sigma$. However, in the opposite limit, $\w>0$ and $T=0$, the scattering
 rate
 is small compared to $\w$ allowing for a straightforward perturbative expansion in the small parameter
 $\Gamma/\w$. This can be done most conveniently  \cite{memory}
 by first multiplying Eq. (\ref{sig}) with $\w^2$ to cancel the
$\delta$-function at $\w=0$.
\begin{equation}\label{sigG}
\w^2 \RE\, \sigma(\w)=  \frac{\IM \left \langle\!\left\langle
\partial_t J,
\partial_t J \right \rangle \!\right\rangle_\w}{\w}=\w^2 \, \RE\, \frac{\chi }{\Gamma(\w)-i
\omega} \approx \chi \RE\,\Gamma(\w)
\end{equation}
 As $\partial_t J$ is already
linear in the interactions (see below), it is sufficient to leading
order to evaluate the correlation function in (\ref{sigG}) to zeroth
order in the couplings. We will use this approximation only at
$T=0$. At any {\em finite} temperature, the scattering rate
$\Gamma(\w\to 0)$ is finite and therefore the method described above
will break down for $\w \to 0$  but remains valid at higher
frequencies where $|\Gamma(\w)|\ll \w$. The drastic simplification
arising for $|\Gamma(\w)|\ll \w$ is also obvious from
Eq.~(\ref{sigma}): while one has to invert the infinite-dimensional
matrix $M$ to obtain the dc-conductivity, one can approximate $\RE
(M-i \w)^{-1}\approx M/\w^2$ if $\w$ is larger than the largest
eigenvalue of $M$.

It is worthwhile to discuss the range of validity of
Eq.~(\ref{sigG}). Quantitatively, an evaluation of $\left
\langle\!\left\langle
\partial_t J,
\partial_t J \right \rangle \!\right\rangle_\w$ to lowest order in
the interactions will only be justified if interactions are weak.
Qualitatively, the lowest order expression does, however, correctly
describe the dynamics of 2-particle collisions. In a Fermi liquid,
$\RE \Gamma(\w)$ will be of order $\w^2$ or smaller due to Pauli
blocking (see below) and therefore smaller than $\w$ in a wide
frequency range even if interactions are strong. $\IM
\Gamma(\w)\approx c \w $ is linear in $\w$ for small frequencies
where $c$ is of order $1$ for strong interactions. While this factor
can lead to strong renormalizations of the optical mass and
therefore of prefactors, such a correction will not change the
power-law dependence of $\RE \sigma(\w)$ for small frequencies
discussed below.

The current of a one-band model,  $\vec{J}=\sum_{\vec{k}\sigma}
\vec{v}_{\vec{k}} c_{\vec{k}\sigma}^\dagger c_{\vec{k}\sigma}$,
decays in the presence of interactions
\begin{equation}
\partial_t \vec{J}=\mi \sum_{\vec{k}
\vec{k}'\vec{q}\sigma\sigma'}U_\vec{q}
(\vec{v}_{\vec{k}}+\vec{v}_{\vec{k}'}-\vec{v}_{\vec{k}+\vec{q}}-
\vec{v}_{\vec{k}'-\vec{q}})c_{\vec{k}\sigma}^\dagger
c_{\vec{k}+\vec{q},\sigma}c_{\vec{k}'\sigma'}^\dagger
c_{\vec{k}'-\vec{q},\sigma'} \end{equation} where we assumed a
density--density interaction with momentum dependence $U_\vec{q}$
(we use $U_\vec{q}\approx U=const.$ for convenience in the
following). $\partial_t J$ is proportional to the difference of
incoming and outgoing velocities. This difference vanishes in a
Galilean invariant system where $v_{\bm k}=\bm k/m$ but is finite in
a lattice model.

Using Eq.~(\ref{sigG}) we obtain for $T=0$ and $\w>0$ to leading
order in $U$
\begin{multline}\label{sigFL}
\RE\,\sigma(\omega>0) \approx \frac{4 \pi U^2}{\w^3}
\sum_{1234\vecG}
f_1 f_2 (1-f_3)(1-f_4)\\
  \times   (v_4^x+v_3^x-v_2^x-v_1^x)^2 \, \delta_{1+2,3+4+\vecG}
  \left[\delta\!\left(\omega- (\eps_4+\eps_3-\eps_2-\eps_1)\right)-(\w
\leftrightarrow -\w)\right]
\end{multline}
where $1,...,4$ denote the momenta $\vec{k}_1,...,\vec{k}_4$ in the
first  Brillouin zone, $f_i=f(\epsilon_i)=f(\epsilon_{\vec{k}_i})$
are Fermi functions and momentum is conserved modulo reciprocal
lattice vectors $\vec{G}$.

To perform the momentum integrals it is useful to split $\vec{k}_i$
into a component perpendicular to the Fermi surface and an angular
integration parallel to it. For small $\w$ only a thin shell of
width $\w/v_F$ contributes for each of the three relevant momentum
integrations perpendicular to the Fermi surface (the 4th integral
takes care of the $\delta$-function arising from energy conservation
and the angular integrations are used to fulfill momentum
conservation). If we neglect the velocity prefactors, this leads to
an $\w^3$ dependence of the integrals which cancels the $1/\w^3$
prefactor.

The velocity prefactors $(v_4^x+v_3^x-v_2^x-v_1^x)^2$ can be
approximated by constants of the order of $v_F^2$, if the Fermi
surface is sufficiently large such that Umklapp scattering processes
can take place. In the presence of Umklapp, one therefore obtains
the well-known result that
\begin{equation}\label{flU}
\RE\,\sigma(\omega>0) \sim \frac{U^2 k_F^5}{v_F^4} \approx const.,
\qquad \Gamma(\w)\propto \w^2
\end{equation}
Note that a constant ``incoherent background'' corresponds according
to Eq.~(\ref{sigG}) to a scattering rate $\Gamma(\w) \propto \w^2$
characteristic of a Fermi liquid with Umklapp scattering in 2 or 3
dimensions.

More interesting is the corresponding result for small Fermi
surfaces ($k_F<G/4$) where Umklapp scattering at the Fermi surface
is not possible and one obtains rather different results in two and
three dimensions. We first consider two dimensions and study the
size of $(v_4^x+v_3^x-v_2^x-v_1^x)^2$ in the limit $\w \to 0$. For a
simple Fermi surface (without pockets etc.), momentum conservation
in the limit $\w\to 0$ can only be fulfilled by combining opposite
momenta, i.e. by choosing $\vec{k}_1=-\vec{k}_2$ and
$\vec{k}_3=-\vec{k}_4$ (or $\vec{k}_1=\vec{k}_{3/4}$ and
$\vec{k}_2=\vec{k}_{4/3}$). As $\bm v_{-\bm k}=-\bm v_{\bm k}$ the
prefactor $(v_4^x+v_3^x-v_2^x-v_1^x)^2$ vanishes linearly in $\w$
for $\w \to 0$ and one obtains from power counting
\begin{equation}
\RE\,\sigma(\omega>0) \propto \w^2, \qquad \Gamma(\w)\propto \w^4
\label{cond2d}
\end{equation}
 for a small Fermi surface in $d=2$.

Momentum conservation is a much less restrictive condition in three
dimensions. For small Fermi surface in $d=3$, momentum conservation
on the Fermi surface does {\em not} require that the relevant
moments are located opposite to each other. For any given momenta
$\bm k_1$ and $\bm k_2$, there is a one-dimensional manifold of
momenta $\bm k_3$ and $\bm k_4$ on the Fermi surface with $\bm
k_1+\bm k_2=\bm k_3+\bm k_4$. For a generic (i.e. non-spherical)
Fermi surface, the prefactor $(v_4^x+v_3^x-v_2^x-v_1^x)^2\sim v_F^2
(k_F a)^4$ will therefore be finite. The suppression by the factor
$(k_F a)^4$, where $a$ is the lattice spacing, describes that the
Fermi surface becomes more and more spherical for small $k_F$ [here
we assumed $\epsilon_{-\bm k}=\epsilon_{\bm k}$ such that leading
corrections to the $k^2/(2 m^*)$ dispersion arise to order $k^4/(a^2
m^*)$]. The finite prefactor implies that
\begin{equation}
\RE\,\sigma(\omega>0)\sim \frac{U^2 k_F^5}{v_F^4} (k_F a)^4 \approx
const., \qquad \Gamma(\w)\propto \w^2 \label{cond3d}
\end{equation}
for a small Fermi surface in $d=3$. Even in the {\em absence} of
Umklapp processes the scattering rate varies as
$\Gamma(T=0,\w)\propto \w^2$! It is, however, suppressed by the
small factor $(k_F a)^4$ compared to a single-particle relaxation
rate. In a Galilean invariant system, this factor vanishes.

The frequency dependence of the scattering rate has to be contrasted
with the temperature dependence discussed in section \ref{sec2}
where we found that $\Gamma(\w=0,T) \sim T^{2 N-2}$ is dominated by
$N$-particle scattering processes with $N>2$ for a small Fermi
surface.

\section{Conclusions}

In a generic clean metal, the frequency and temperature dependence
of the optical conductivity arise from rather different physical
processes and have in general very little in common. This can be
seen most drastically in systems where momentum conservation (or its
generalization the pseudo-momentum conservation) dominates
transport. In such a situation, we found  $\Gamma(\w,T=0)\sim \w^2$
while $\Gamma(\w=0,T) \sim T^{2N-2}$ where $N>2$ depends on the size
of the Fermi surface (or its curvature in the case of a quasi
one-dimensional metal). This disparate behavior can be traced back
to the simple fact that momentum and current are two different
operators in a lattice system. The response at finite frequencies
depends on the relaxation rate  of the current operator. We have
shown that in a three-dimensional metal current can relax
efficiently even in the absence of Umklapp scattering as a typical
low-energy scattering process of two particles changes the current
by a finite amount $(v_4^x+v_3^x-v_2^x-v_1^x)^2\sim v_F^2 (k_F
a)^4$. The same type of process, however, does not relax the
momentum. We have argued that this implies that a finite fraction of
the current -- the fraction which is 'parallel' to the momentum --
does not relax by this process. The consequence is that such a
process is not effective in determining the $\w=0$ relaxation rate
at finite temperatures. The $T$ dependence of the dc conductivity is
given by the relaxation rate of the (pseudo-) momentum and therefore
a completely different set of processes.

As the $\w$ and $T$ dependence are completely independent for small
Fermi surfaces or quasi one-dimensional metals, it is obvious that
also in the {\em presence} of Umklapp scattering, when
$\Gamma(\w,T)\approx a (k_B T)^2+ b (\hbar \w)^2$ there is in
general no simple relation between the constants $a$ and $b$. We
emphasize this fact since in the experimental literature such a
relation has sometimes been claimed to exist
\cite{dressel,reviews,exp1} but is actually not observed
\cite{exp1,scheffler}.

When discussing the role of momentum conservation on the one hand
and Umklapp scattering on the other hand it is important to keep in
mind that in general the concept of momentum is not uniquely well
defined in a lattice. We have emphasized that in general one has to
search for a slow mode. In the case of strongly anisotropic Fermi
liquids (or Luttinger liquids  \cite{roschPRL}, spin chains and
ladders  \cite{shimshoni} or superconductors  \cite{dWave} with
point nodes) this has motivated us to consider so-called
pseudo-momenta, see Eq.~(\ref{pseudo}), which can be identified with
the momenta of effective low-energy field theories  \cite{roschPRL}.

More than a century after the pioneering work of Drude the
measurement and calculation of the optical conductivity remains an
interesting and challenging area of condensed matter physics. Open
questions include for example the transport close to
quantum-critical points where the role of momentum conservation and
Umklapp scattering deserves further investigations. But even in
simple Fermi liquids it is interesting -- both experimentally and
theoretically -- to study the precise shape of the Drude peak in
regimes where impurity scattering can be neglected, especially in
quasi one-dimensional metals \cite{dresselBech}.

\begin{acknowledgement} I would like to thank  N. Andrei and P. W\"olfle for
helpful discussions at various stages of this work and the DFG for
financial support under SFB 608.
\end{acknowledgement}


\begin{thebibliography}{10}
\bibitem{new} In the case of a small Fermi surface where momentum
  conservation dominates, we  expect this to be a long-existing result but
  we were not able to find a related reference.
\bibitem{remark}  Additional
logarithmic temperature dependences of the scattering vertices are
expected even in the Fermi-liquid regime as it is well known from
Fermi liquid theory.
\end{thebibliography}
\end{document}